\documentclass[
aps,
prfluids,
superscriptaddress,
amsmath,
amssymb,
longbibliography,
]{revtex4-2}

\usepackage{graphicx}
\usepackage{dcolumn}
\usepackage{bm}

\begin{document}

\title{Ray-tracing image simulations of transparent objects with complex shape and inhomogeneous refractive index}

\author{Armin~Kalita}
    \affiliation{Department of Physics, Rutgers University-Newark, Newark, NJ 07102, USA}
\author{Bryan~Oller}
    \affiliation{Department of Physics, Rutgers University-Newark, Newark, NJ 07102, USA}
\author{Thomas~Paula}
    \affiliation{Chair of Aerodynamics and Fluid Mechanics, TUM School of Engineering and Design, Technical University of Munich, 85748 Garching bei M\"{u}nchen, Germany}
\author{Alexander~Bu{\ss}mann}
    \affiliation{Chair of Aerodynamics and Fluid Mechanics, TUM School of Engineering and Design, Technical University of Munich, 85748 Garching bei M\"{u}nchen, Germany}
\author{Sebastian~Marte}
    \affiliation{Department of Physics, Rutgers University-Newark, Newark, NJ 07102, USA}
\author{Gabriel~Blaj}
    \affiliation{Technology Innovation Directorate, SLAC National Accelerator Laboratory, Menlo Park, CA 94025, USA}
\author{Raymond~G.~Sierra}
    \affiliation{Linac Coherent Light Source, SLAC National Accelerator Laboratory, Menlo Park, CA 94025, USA}
\author{Sandra~Mous}
    \affiliation{Linac Coherent Light Source, SLAC National Accelerator Laboratory, Menlo Park, CA 94025, USA}
\author{Kirk~A.~Larsen}
    \affiliation{Linac Coherent Light Source, SLAC National Accelerator Laboratory, Menlo Park, CA 94025, USA}
\author{Xinxin~Cheng}
    \affiliation{Linac Coherent Light Source, SLAC National Accelerator Laboratory, Menlo Park, CA 94025, USA}
\author{Matt~J.~Hayes}
    \affiliation{Linac Coherent Light Source, SLAC National Accelerator Laboratory, Menlo Park, CA 94025, USA}
\author{Kelsey~Banta}
    \affiliation{Linac Coherent Light Source, SLAC National Accelerator Laboratory, Menlo Park, CA 94025, USA}
\author{Stella~Lisova}
    \affiliation{Linac Coherent Light Source, SLAC National Accelerator Laboratory, Menlo Park, CA 94025, USA}
\author{Peter~Nguyen}
    \affiliation{Linac Coherent Light Source, SLAC National Accelerator Laboratory, Menlo Park, CA 94025, USA}
\author{Serge~A.~H.~Guillet}
    \affiliation{Linac Coherent Light Source, SLAC National Accelerator Laboratory, Menlo Park, CA 94025, USA}
\author{Divya~Thanasekaran}
    \affiliation{Linac Coherent Light Source, SLAC National Accelerator Laboratory, Menlo Park, CA 94025, USA}
\author{Silke~Nelson}
    \affiliation{Linac Coherent Light Source, SLAC National Accelerator Laboratory, Menlo Park, CA 94025, USA}
\author{Mengning~Liang}
    \affiliation{Linac Coherent Light Source, SLAC National Accelerator Laboratory, Menlo Park, CA 94025, USA}

\author{Stefan~Adami}
    \affiliation{Chair of Aerodynamics and Fluid Mechanics, TUM School of Engineering and Design, Technical University of Munich, 85748 Garching bei M\"{u}nchen, Germany}
\author{Nikolaus~A.~Adams}
    \affiliation{Chair of Aerodynamics and Fluid Mechanics, TUM School of Engineering and Design, Technical University of Munich, 85748 Garching bei M\"{u}nchen, Germany}
\author{Claudiu~A.~Stan}
    \email{Email: claudiu.stan@rutgers.edu}
    \affiliation{Department of Physics, Rutgers University-Newark, Newark, NJ 07102, USA}


\begin{abstract}
Optical images of transparent three-dimensional objects can be different from a replica of the object’s cross section in the image plane, due to refraction at the surface or in the body of the object. Simulations of the object’s image are thus needed for the visualization and validation of physical models. We report ray tracing image simulations that achieved high physical fidelity, reproducing optical behaviors and image features not rendered in previous studies. We replicated brightfield microscopy images of drops with complex shapes, and images of pressure and shock waves traveling inside them. For high physical fidelity, the simulations must replicate the spatial and angular distribution of illumination rays, and both the experiment and the simulation must be designed for accurate optical modeling. The simulations are highly sensitive to the properties of the drops and can be used to diagnose and refine fluid dynamics models. The simulated images can also be optimized to extract multiple 3D properties from experimental images. Compared to specialized single-shot 3D imaging methods, this approach has the advantage that it preserves the experimental simplicity, the high resolution, and the visual interpretability characteristic to basic optical imaging. The techniques introduced here are directly applicable to optical microscopy, so they can be used in other fields, such as microfluidics and biology, to expand the type and the accuracy of three-dimensional information that can be extracted from basic optical images.
\end{abstract}

\maketitle

\section{\label{sec1}Introduction}

The optical images of transparent objects are often distorted due to the refraction of light. The entire volume of the object is relevant to image formation, because rays refract at interfaces and curve through refractive index gradients. This volumetric distortion of optical images restricts the information that can be extracted from them, which  is limiting when the objects experience rapid dynamics that preclude three-dimensional (3D) imaging techniques such as tomography and confocal microscopy, or when the degradation of resolution with single-shot 3D imaging \cite{levoy3D} is unacceptable.

The optical distortions generated by transparent objects encode 3D information about the object in the image, but this information cannot be retrieved because the encoding is too complex. This contrasts with techniques such as light field microscopy \cite{levoy3D} and holographic microscopy \cite{garcia, martin}, which encode 3D information such that it can be retrieved via reconstruction or fitting. In principle, the 3D information provided by optical distortions might be decoded via iterative procedures that model images of trial objects to replicate the experimental image. In practice, this iterative approach was not possible because it requires image simulations with a higher fidelity than achieved to date.

Ray tracing, a basic technique for understanding image formation, can be used to simulate images by tracing a large number of rays to a virtual detector \cite{glassner}. While one of its best known applications is generating photorealistic images in computer graphics, some of the most sophisticated ray-tracing image simulations for real systems are related to fluid dynamics \cite{brownlee, luthman, koch}. Fluid dynamics presents challenging imaging problems, such as 3D systems with complex and dynamic boundaries, or nonuniform refractive index distributions. The latter are caused by the dependence of refractive index on the density, which can vary smoothly due to pressure waves, sharply due to shock waves, or jump at phase interfaces. 

Ray-tracing simulations for fluid dynamics focused primarily on the visualization of the dynamics predicted by computational fluid dynamics (CFD) simulations, because simply plotting CFD-calculated object shapes and density distributions may not replicate the experimental images due to optical distortions \cite{koch}. This problem led to the development of analytical techniques for simulating experimental images \cite{yates, settles}, such as plotting integrated density gradients to simulate Schlieren images \cite{yates}. Numerical ray tracing improved these early techniques. For example, tracing rays through density gradients in the object produced more realistic Schlieren images \cite{brownlee}.

The visualization of CFD results was further improved by tracing the rays through the entire optical system, from the illumination source to the camera. This technique was applied to Schlieren and shadowgraph imaging of shocks and rarefaction waves in a gas \cite{luthman}, and to shadowgraph imaging of collapsing bubbles in a liquid \cite{koch}. The experimental and simulated images were similar, but the simulated images had limited physical fidelity,  being for example much sharper than the experimental ones \cite{koch}.

Prior work shows that ray-tracing image simulations are challenging for 3D objects in brightfield imaging \cite{luthman, koch}. Interferometric and holographic microscopy \cite{garcia, martin} can provide 3D information from single-shot images, but they typically use collimated beams that limit optical resolution and the ability to image inside strongly refracting objects, the recorded images are not visually interpretable, and the reconstructed images often have artifacts. This makes them less suitable for imaging faint localized features caused by shock and pressure waves.

The difficulties of interpreting optical images can be mitigated by using ultrafast X-ray phase contrast imaging (XPCI) \cite{schropp, vassholz, hodge, makarov}. Materials have an X-ray refractive index close to 1, thus XPCI images are less affected by refractive distortions but may have fainter image features. So far, ultrafast XPCI has been used for high contrast features such as phase boundaries and shock waves. Hydrodynamic simulations for these studies \cite{schropp, vassholz, hodge, makarov} generally agreed with experimental XPCI data for shock waves with amplitudes from several to hundreds of GPa, but simulations of XPCI images had low fidelity, in part due to the difficulty of calibrating the X-ray illumination \cite{hodge}.

Here, we report ray-tracing simulations for optical shadowgraph imaging that replicated images of spherical and non-spherical drops, in focus and out of focus, with significantly better fidelity than in previous shadowgraph simulations \cite{koch}. The simulations also replicated images of shock waves and of smooth pressure waves with amplitudes on the order of 100 MPa. The improvement of fidelity enabled new qualitative and quantitative applications of ray-traced image simulations, including testing and refining independent physical models, and measuring 3D parameters such as axial position and orientation.

\section{\label{sec2}Results}
\subsection{\label{sec2A}Experimental setup and the design of the ray-tracing model}

The limited fidelity of previous ray-tracing simulations is difficult to diagnose due to the high complexity of ray tracing models. In principle, this could be due to a fundamental limitation of ray-tracing such as ignoring diffraction, but here we were able to improve the fidelity without considering diffraction. For simulations that model all elements of the imaging system \cite{luthman,koch}, the fidelity might be improved by optimizing the parameters of the optical components, but we found that tuning parameters to minimize the difference between simulated and experimental images did not solve the fidelity problems.

Eventually, we found that the key limitation to fidelity was the incomplete treatment of the illumination. For a first demonstration of high fidelity, it was important to minimize the non-ideal features of the experimental system, such as nonuniform illumination or optical aberrations, because they require a more complicated ray tracing model where it is harder to discern what factors affect the fidelity. Also, given the complexity of the models and the possibility of hidden limitations in the ray tracing software, it was critical to test the simulations thoroughly.

We imaged three types of free-falling water drops in free space: spherical drops, non-spherical merging drops, and drops containing pressure waves induced by laser ablation. The spherical drops were used as an object with known geometry. The other drops were used to test all the interactions of rays with an optically isotropic transparent object: refraction and partial reflection at interfaces, and ray bending in inhomogeneous refractive index regions. The merging drops and the ablated drops are evolving systems from which we extracted tens of distinct objects with (i) convex, concave, and saddle surfaces, and (ii) positive and negative refractive index modulations in regions with different curvatures. This set is physically comprehensive in terms of the possible optical phenomena, and probes more conditions than previously \cite{luthman, koch}.

\begin{figure}
\includegraphics{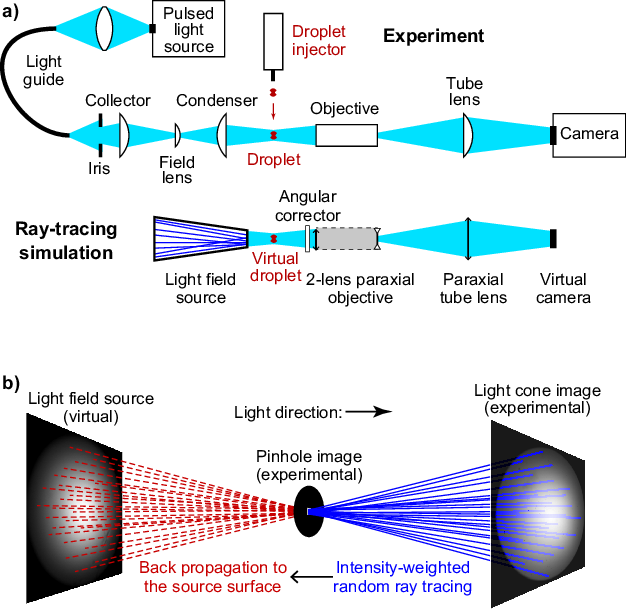}
\caption{\label{fig1}The design of the ray-tracing simulation. (a) The illumination source and optics are modeled with a virtual light field source that replicates the spatial and angular distribution of rays in the image plane, and the imaging optics are modeled with ideal lenses. The angular corrector models the angle-dependent reflective losses in the real setup. (b) Generation of the light field source. Rays are first traced between an image of a pinhole at the object location and the image of the light cone emerging from the pinhole, then back traced to the light field source. }
\end{figure}

Figure~\ref{fig1}(a)  illustrates the imaging system, which is a horizontal microscope. A vibrating nozzle generated drops with velocities near 10 m/s, and individual drops were imaged using light pulses as they passed through the imaged region. The merging drops were injected as groups of three drops with slightly different velocities (±0.2 m/s) that caught up and collided, while the ablated drops were exposed to an X-ray laser pulse before imaging (see the Supplemental Material for details \cite{suppmat}).

To enable accurate image simulations, the experimental setup must generate high quality images, and its components must be chosen such that they can be modeled accurately. Since modeling more components of the optical system increases the ray tracing accuracy \cite{luthman, koch}, the simulation accuracy is maximized when each component is modeled. If this strategy is chosen, reflective objectives are easiest to model because they contain only two mirrors. Nevertheless, the focus field of the reflective objectives we tested was not flat enough for our experiments.  Corrected refractive objectives provide sufficiently flat fields, but they have proprietary designs, so in practice they cannot be modeled exactly. We chose a refractive objective that provided the required image quality and could be modeled with ideal lenses (Mitutoyo M Plan Apo 20×, NA = 0.42, infinity corrected).

The objective was paired with an f = 200 mm tube lens and a high-speed camera (Vision Research Miro M340). The optical resolution at 462 nm, with the objective aperture fully illuminated, was 0.7 µm. The illumination source was either a pulsed LED (HardSoft IL-106B, 462 nm, 100 ns pulses) or a femtosecond laser (frequency-doubled Ti:sapphire laser system \cite{liang}, central wavelength 405 nm, 30 fs pulses). The pulses were shorter than the camera shutter and defined the exposure time. The light pulses from the source were projected into a liquid light guide to homogenize the beam (Thorlabs LLG3-6H, 3 mm diameter, 1.8 m long), and then imaged onto the sample using an f = $+$42.5 mm collector lens, an f = $+$150 mm field lens, and an f = $+$25 mm condenser lens. The NA of the illumination was adjusted with an iris at the collector. This setup produced a circular illuminated spot with a diameter of 2.7 mm and less than 2\% intensity variation where the drops were imaged ($<$200 $\mu$m from the center of the spot).

The corresponding ray-tracing setup (Fig.~\ref{fig1}(a)) modelled the entire illumination with a virtual illumination source that replicated, across the object plane, the experimental light intensity distribution and the experimental angular distribution of incoming rays. This information constitutes the 4D light field \cite{levoyLF} of the illumination beam, thus our virtual illumination source is a light field source. The experimental light field was measured by placing a pinhole at the object location and projecting the emerging light cone on a camera. The virtual light field source was then generated as illustrated in Fig.~\ref{fig1}(b), by tracing rays between the images of the pinhole and of the light cone, followed by back tracing these rays to the light field source, which is a ray-emitting surface located before the object. This experimentally-matched light field source was a key factor in achieving a good match between experimental and simulated images, because the drop images are highly sensitive to the light field \cite{suppmat}.

The imaging optics were modelled with paraxial lens combinations and a virtual camera, with parameters replicating the physical ones. Since the objective had a working distance of 20 mm, longer than its 10-mm focal length, we modelled it with two paraxial lenses located at the ends of the physical objective. The objective lens parameters were calculated from the focal length, the working distance, and the pupil sizes of the physical objective \cite{suppmat}. To account for angle-dependent reflection losses in real lenses, the model includes an angular corrector placed in front of the objective. The corrector was made of a central circle surrounded by eight rings with experimentally calibrated transmission factors \cite{suppmat}.

The model was implemented in Ansys Zemax OpticStudio (ANSYS, Inc.). Zemax is mainly used for optical design, but it is also well suited for ray-tracing image simulations. Compared to the home-built and open-source software used previously \cite{luthman,koch}, Zemax has a detailed manual, models many optical elements, and has an intuitive design interface, all which allow rapid and precise model design and testing. Its programming interface enabled us to run complex simulations efficiently via MATLAB (Mathworks, Inc.).

The images were simulated by tracing rays from the source to the camera, using the non-sequential mode to model partial reflections at interfaces. This method builds the image stochastically by accumulating ray hits in individual pixels, similar to how photons generate images in real cameras. The number of ray hits was approximately equal to the number of photons recorded in experimental images, to match the photon shot noise. We estimated the experimental number of photons per pixel as the camera signal on a 0 to 1 scale multiplied by its electron well capacity.

\subsection{\label{sec2B}High fidelity image simulations of multiple classes of test objects}

\begin{figure}
\includegraphics{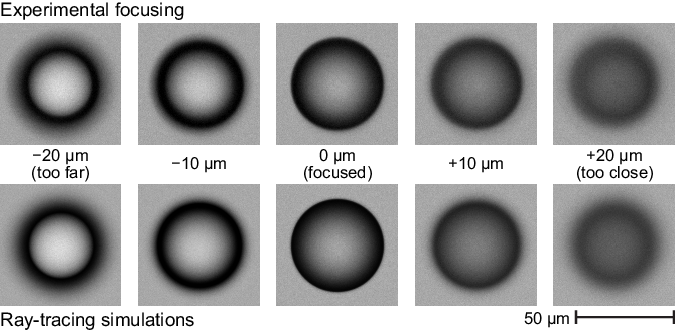}
\caption{\label{fig2}Reproduction of a focus scan. Experimental and simulated images of spherical water drops are compared as the objective was focused. The simulations reproduce the intensity distribution within the drop image under all conditions.}
\end{figure}

Figure~\ref{fig2} shows a comparison of experimental and simulated images of 46-µm diameter spherical water drops, imaged as the objective was scanned $\pm$20 µm across the focus. The iris at the collector was fully open and the illumination rays had a maximum angle of $27^{\circ}$, fully filling the numerical aperture of the objective. The ray-tracing simulations reproduced accurately the experimental images over a scan distance one order of magnitude larger than the depth of focus ($\lambda$/NA$^2$ = 2.6 $\mu$m). Additional simulations for up to $\pm$40 µm defocusing are shown in the Supplemental Material \cite{suppmat}; differences from the experiment are evident only at $-$40 µm.

The simulated images also illustrate the limitations of basic ray racing. Simulated images were sharper than the experimental ones because they did not consider diffraction or the vertical motion of the drops, and less noisy because they did not include the electronic noise of the camera.  The simulations can be modified to model motion, diffraction, and camera noise, and an example is given in the Supplemental Materials \cite{suppmat}. However, these refinements were not needed to demonstrate the physical fidelity of  ray-tracing simulations, so we did not use them in the main text.

The focus scan of a spherical drop is a sensitive test of the simulations, because slight imperfections of the model may be evident only at some focusing positions, and because it clarifies if the simulated images are unrealistically sharp \cite{settles, koch}. Focus scans for variations of the model showed that a single-lens objective model is equivalent to the two-lens model, while using a different light field, ignoring partial reflections, or omitting the angular corrector degrades the match between simulation and experiment \cite{suppmat}.

For test objects more complicated than a sphere with constant refractive index, we built object models using experimental data or CFD simulations. In practice, multiple models were possible within the experimental accuracy, and one had to be chosen. For example, the time delays of CFD-generated objects were selected from delays in the range of experimental uncertainty to produce the closest match to experiment. This approach is valid since our drops had deterministic evolution with unique shapes or pressure wave locations.

\begin{figure}
\includegraphics{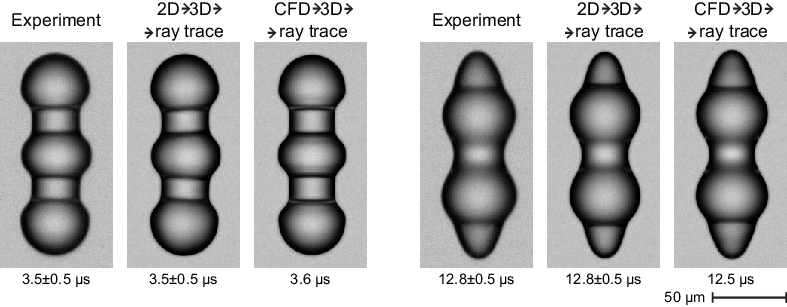}
\caption{\label{fig3}Reproduction of images of objects with complex shapes. The figure shows images of a merging drop formed by the collision of three spherical drops at time zero, at two different time delays. The experimental images are compared with simulations of 3D drops generated by revolving the experimental 2D outline of the drop image, and with simulations of 3D drops computed using CFD.}
\end{figure}

The ray tracing model reproduces accurately images of drops with complex shapes. Figure~\ref{fig3} shows images of aspherical drops, which were generated by merging three spherical drops. The aspherical drops assume multiple shapes as they relax, and Fig.~\ref{fig3} shows two of the observed shapes. In this case the drops cannot be modeled with a simple geometrical shape, and we used two methods to generate a three-dimensional model. The first method revolved the 2D outline of the drop from the experimental image, which generated a 3D drop. The second method calculated the 3D shape of the drop using the compressible multiphase flow solver ALPACA \cite{adams}. Detailed descriptions of both methods are available in the Supplemental Material \cite{suppmat}, which also contains movies showing additional delays, for drops merging in focus and for drops merging while traveling across the focal plane.

The refraction of light by the drops in Fig.~\ref{fig3} leads to a pattern of light intensity across the images of the drops. These patterns are reproduced nearly identically in simulations, including subtly brighter or darker regions. Overall, the simulated images replicated the experimental ones with higher fidelity than in previous studies \cite{koch}. Slight differences between experiment and simulations exist because the simulated drops did not replicate perfectly the shape of the experimental drops, which were not precisely axisymmetric.

A stringent test of the ray-tracing simulations is replicating images of objects with smooth spatial variations in their refractive index. In this case, the image cannot be understood as a mapping of the 3D object space onto the 2D image space, because the object does not have sharp identifiable features. As a test of this case, we simulated images of water drops ablated by an X-ray laser pulse. The drops were imaged optically after they were exposed to a femtosecond X-ray laser pulse passing approximately through the center of the drop. The X rays heated isochorically a thin filament of water to high temperatures and pressures, which launched a cylindrical shock wave. The shock reflected at the surface of the drop, generating a converging negative pressure wave \cite{stanNP}.

The experimental ablation images were recorded at the CXI instrument of the LCLS X-ray laser facility \cite{liang, emma}, using 43 $\mu$m diameter water drops. The imaging system was identical to the one used for the spherical and the merging drops, except for (i) using 30 fs laser pulses for illumination, and (ii) reducing the maximum illumination angle to approximately $23^{\circ}$ to enhance the contrast of waves. The X-ray pulse energy was below the threshold for generating secondary cavitation (spallation) in the drops \cite{stanNP}, which allowed us to observe the negative pressure wave converging to a focus and then expanding.

The simulated images used virtual drops with properties calculated using CFD simulations performed in ALPACA \cite{adams}.  These simulations were based on a CFD model that reproduced the X-ray ablation dynamics for conditions different from our experiment \cite{paula19}. We adjusted the initial conditions of the CFD model to replicate the experimental motion of the drop surface during shock reflection \cite{stanNP, stanSACLA, suppmat}. We also optimized the NASG equation of state \cite{lemetayer} used in the CFD simulations to match the adiabatic and shock compression paths predicted by the IAPWS-95 equation of state \cite{wagner, suppmat}. The results of the CFD simulations provided the 3D shape of the drops, and their internal pressure and density distributions. The refractive index distribution was calculated from the density distribution, the temperature of the drop, and a light wavelength of 400 nm \cite{harvey}.

\begin{figure}
\includegraphics{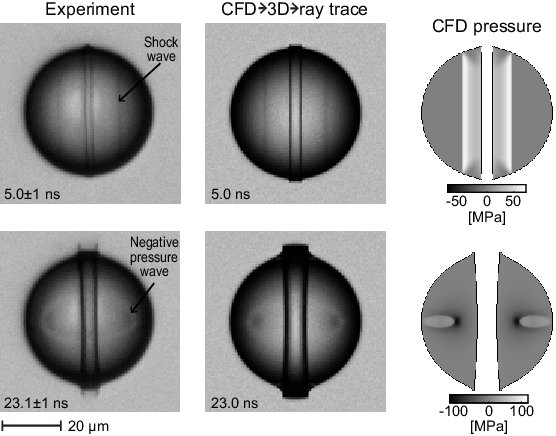}
\caption{\label{fig4}Reproduction of images of laser-ablated drops with internal pressure waves. The image simulations replicate the width and the intensity of shock waves and of negative pressure waves, and are qualitatively different from the corresponding 2D pressure distributions in the image plane. The central vertical feature is an ablation hole, which has a different size in simulations because it is approximated in the CFD model.}
\end{figure}

Figure~\ref{fig4} shows experimental and simulated images of an expanding shock wave with an CFD-calculated amplitude of 70 MPa, and of a converging negative pressure wave with an amplitude of $-$110 MPa. The simulations replicate the width and intensity of the shock better than it was previously achieved with either brightfield or Schlieren imaging \cite{luthman}. The image of the negative pressure wave, which to our knowledge is the first replication of a smooth pressure wave in brightfield images, matched the experimental aspect ratio and the pattern of dark and bright modulations. Comparisons at additional time delays are available as movies in the Supplemental Material \cite{suppmat}, for the experiment shown in Fig.~\ref{fig4} and for another one with a lower energy X-ray pulse, which generated visible pressure waves without producing ablation holes. The movies show that our simulations replicate a counterintuitive optical effect—the shocks become more visible as they propagate, despite becoming weaker.

Due to experimental and CFD limitations, the overall match of images for ablated drops is worse than for the merging drops. Experimentally, (i) we had to use a surrogate light field that did not reproduce a bright spot in the experimental images, and (ii) the image of the negative pressure wave was affected by misalignments between the drop and the X-ray pulse (see Sec.~\ref{sec3B}). Modelling-wise, the CFD simulation does not treat accurately the initial formation of the ablation core (see Sec.~\ref{sec3A}), so the images of the core are different. Even with these limitations, the simulated images had a higher degree of fidelity than in XPCI studies \cite{hodge}.

\subsection{\label{sec2C}Overall image similarity}

In addition to the higher fidelity, the simulations shown in Figs.~\ref{fig2}–\ref{fig4} are arguably more similar to the experimental images than in previous simulations of spherical drops \cite{koch}, bubbles with complex shapes \cite{koch}, and shock or pressure waves \cite{luthman,hodge}. The overall image similarity can be judged quickly via visual inspection, but it does not imply fidelity in the reproduction of faint image features due to pressure waves. Thus, our criterion for developing these ray-tracing simulations was to achieve high fidelity. Nevertheless, the quantification of image similarity is important for applications.

Defining a numerical measure that captures all the image similarities that can be discerned by the human vision system (HVS) is a difficult and longstanding problem in imaging. The HVS is very sophisticated system whose discriminating capabilities could not be fully replicated yet by simple numerical measures \cite{wangSSIM}. A basic measure of image similarity is the mean squared error (MSE), or the mean squared difference between images. MSE is objective and simple, but does not capture well the perceived similarity between images \cite{wangSSIM}. In our case, using MSE is additionally problematic because it has a strong dependence on the image noise, which may mask differences between weak image features due to pressure waves.

We explored using MSE and three structural similarity measures designed to mimic the discriminating capabilities of the HVS: the basic structural similarity index (SSIM \cite{wangSSIM}), the multiscale structural similarity index (MSSIM \cite{wangMSSIM}), and the complex wavelet structural similarity index (CWSSIM \cite{sampat}). These similarity measures are numbers with a maximum value of 1 for identical images. MSSIM and CWSSIM are multiscale indices whose maximum possible scale increases with the image size. The maximum scale represents the highest level of detail in image comparisons, so in principle it should offer the most stringent comparison between images. However, we found that at the maximum scale the similarity values saturated at values very close to 1, and were prone to mis-identifying the most similar images from simulation scans. Therefore, we calculated MSSIM and CWSSIM at one unit below the maximum possible scale. To calculate the MSE, SSIM and MSSIM, we used the functions implemented in MATLAB, and for CWSSIM we used the ''cwssim\_index'' function from the MATLAB file exchange \cite{sampat}.

We tested these similarity measures by evaluating the cross-similarity between experimental and simulated images from an extended focus scan \cite{suppmat}. MSE, MSSIM, and CWSSIM produced the physically correct result of generating the highest similarity index or the lowest MSE when experiments and simulations had the same focus position, but the SSIM for the experiment at 0 $\mu$m was unphysically larger with the simulation at $-$5 $\mu$m than the one at 0 $\mu$m. MSE, MSSIM, and CWSSIM can thus be used to determine which image in a simulated focus scan matches best the experiment. They can also be used as error functions for image-based optimization (Sec.~\ref{sec3D}).

Image similarity measures cannot be used for comparisons between ray-tracing models of different imaging systems. For example, Koch \emph{et al.} \cite{koch} reported image similarity values for ray-traced image simulations. They compared experimental and simulated images of spherical drops, obtaining SSIM values ranging from 0.420 to 0.619 and CWSSIM values ranging from 0.352 to 0.721. Our only option for a direct comparison was to compute SSIM values for our spherical drops in Fig.~\ref{fig2}. We could not calculate similarity values for Koch et al. because their images are compressed (which lowers the similarity \cite{wangSSIM}), and we could not calculate an equivalent CWSSIM because their CWSSIM input parameters were not specified. The SSIM values for the images in Fig. 2 range from 0.74 to 0.82 and are higher than in Koch \emph{et al.} Nevertheless, their images had more complicated features which may depress the similarity values, and SSIM failed our focus scan test, so based on the SSIM values it is not possible to decide which set of simulations is better. More generally, image similarity measures cannot provide absolute measurements of similarity because they depend on image noise, size, and composition. In contrast with this difficulty in comparing similarity based on numerical measures, the improvement in fidelity (reproducing the focusing behavior) is unambiguous.

\subsection{\label{sec2D}Computation times and fast noisy simulations}

Simulating images with the same number of ray hits as the number of photons captured in the experimental images is computationally intensive. For example, the droplet merging images were simulated with a virtual sensor size of 200×300 pixels, and the experimental images recorded around 2500 photons per pixel. To replicate the experimental image with the same number of ray hits, 200×300×2500 = 150 million hits are needed. Since only part of the rays from the source arrive at the detector, around 400 million rays were traced from the source. Furthermore, if the object does not have a basic shape, a numerical surface must be used. A complex shape can increase substantially the computation time if the surface data has high resolution. On a 64-core workstation (Lenovo P620 with an AMD Ryzen ThreadRipper PRO 3995WX processor), simulating an image of a spherical drop took approximately 10 minutes, but simulating an image of a merging drop took approximately one hour.

Computation times increased further when tracing rays through drops with nonuniform refractive index distributions caused by pressure waves. To consider the variations in refractive index, the ray path must be calculated in much smaller length steps than for a homogenous object, and this increases the computation time. Additionally, the size of refractive index data impacts the speed. For axisymmetric drops, the refractive index distribution can be efficiently stored in a 2D matrix, and the computation time for laser-ablated drops was comparable to the one for the merging drops. If the drop is not axisymmetric due to misalignments (see Sec.~\ref{sec3B}), the refractive index distribution is stored in a 3D matrix, which leads to substantially longer computation times. The most complex simulations, with 3D refractive index and surface data, were used for Fig.~\ref{fig4} and Supplementary Movie 4, and required around 10 hours of computations per image.

Although the computations are lengthy, ray tracing is parallelizable with a linear speed gain, which allows efficient use of multi-core workstations, computer clusters, or cloud computing. Since Zemax uses CPU computations, another possibility is using GPU computations. GPU ray-tracing at low numerical resolution can be substantially faster than CPU ray-tracing \cite{kimura}, but it might require algorithms that mitigate the numerical noise \cite{wu}. These solutions require additional resources or future developments.

An already available solution is to simulate images with much fewer ray hits than photons. In Figs.~\ref{fig2}–\ref{fig4}, to explore how well ray tracing replicates experimental images, we used simulations with the same number of rays as photons. However, many image features can be resolved with fewer rays. Figure 5 shows image simulations for several images from Figs.~\ref{fig2}–\ref{fig4}, made using down to 1/1024 of the full number of rays. The features of spherical and merging drops degraded substantially between 1/64 and 1/256 rays, while the fainter features caused by shocks and pressure waves degraded substantially between 1/16 and 1/64 rays.

\begin{figure}
\includegraphics{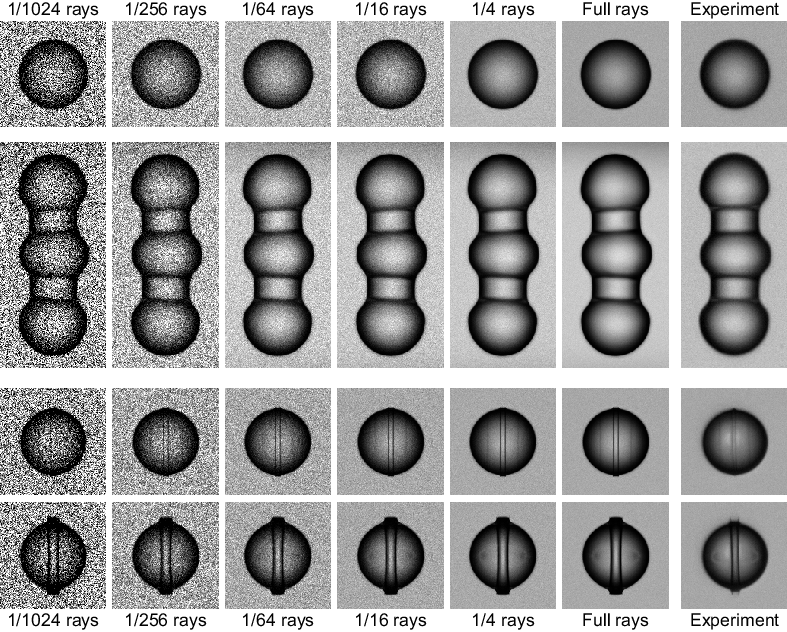}
\caption{\label{fig5}The effect reducing the number of rays, for images from Figs.~\ref{fig2}–\ref{fig4}. Faint wave features remain visible after reducing the number of rays by one order of magnitude, and some details of the drop images remain visible even after reducing the number of rays by three orders of magnitude.}
\end{figure}

The ability to preserve key image features in Fig.~\ref{fig5}, even when the number of rays is decreased by orders of magnitude, is due to the favorable scaling of the ray-tracing shot noise. Since the relative shot noise scales with the inverse square root of the number of rays, reducing the number of rays 100-fold leads to only a 10-fold increase in noise. This high tolerance to noise enables the use of fast noisy simulations not only for qualitative investigations, but also for optimizing 3D object parameters. Section III.D describes the simultaneous optimization of two depth parameters of the merging drops, using images with 1/1024 fewer rays than in Fig.~\ref{fig3}.

\section{\label{sec3}Application examples and discussion}
\subsection{\label{sec3A}Qualitative testing and refinement of physical models}

\begin{figure}
\includegraphics{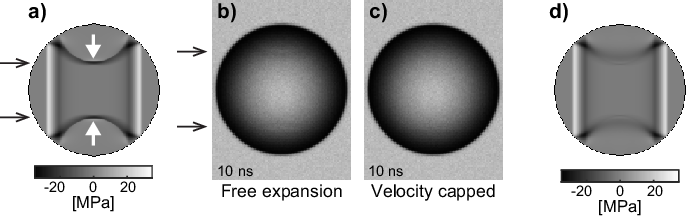}
\caption{\label{fig6}Refinement of the CFD model using image simulations. (a) Pressure distribution from an initial CFD model, and (b) corresponding simulated image. Both display an unphysical pressure wave that was not observed in the experiment. This wave is indicated by white arrows in the CFD plot, and the black arrows indicate the height of the tips of this wave. (c,d) Corresponding results from a velocity-capped CFD model that restricts the expansion of the filament into the surrounding gas. The unphysical pressure wave no longer appears in the image. Differences between the simulated images may only be visible on a computer screen.}
\end{figure}

In initial simulated movies of the weak ablation case without hole formation, we observed an additional pressure wave traveling from the regions where the X-ray pulse intersected the surface of the drop. A plot of the corresponding pressure distribution is shown in Figure~\ref{fig6}(a). Since the amplitude of this additional pressure wave is smaller than the shock and its reflection, which are themselves barely visible in the experiment, it cannot be determined from CFD data if the additional wave would be visible in experiments (and thus real). However, according to the simulated image shown in Fig.~\ref{fig6}(b), this wave would generate visible features not seen in the experiment. Therefore, this wave either does not exist in the experiment, or it has a lower amplitude. In Fig.~\ref{fig6}, because the unphysical wave is difficult to see in single frames, we used a higher filament pressure (275 MPa) than the best-matched pressure (195 MPa) used for Supplementary Movie 3.

This unphysical wave is a rarefaction wave that appeared in the initial CFD simulations because the parts of the X-ray heated filament near the drop surface expanded into space freely and faster than the rest of the filament, which was confined by the surrounding liquid. However, since the filament is hot, its ends will evaporate. Evaporation generates a reactive pressure equal to the vapor pressure \cite{stanRD}. This reactive pressure will restrict the expansion of the filament. Based on the energy of the X-ray pulse, we estimated that the filament becomes superheated, which leads to high vapor pressures, possibly up to the critical pressure of 22 MPa.

To suppress the unphysical wave in simulations, we extended the CFD model with a modified interface condition. By default, at the interface between water and air, we solved a two-phase Riemann problem to obtain the velocity of the interface as well as the numerical evolution terms for both phases \cite{paula23}. The reactive pressure due to evaporation was missing from our model and thus also missing from the Riemann problem. To account for the reactive pressure, we assumed that it suppresses the interface movement at the ends of the filament, and we added an additional interface Riemann problem that constrains the interface velocity to zero and computes the remaining states accordingly. To avoid abrupt changes in the interface conditions, we blended between the constrained Riemann problem and the standard Riemann problem with a Gaussian profile, exp($-$2/(x/5 $\mu$m))$^2$, where x is the distance from the drop axis.

Image simulations based on the velocity-capped CFD model, shown in Fig.~\ref{fig6}(c), no longer displayed the wave. A rarefaction wave caused by the oblique reflection of the shock wave can be observed in the pressure distribution from Fig.~\ref{fig6}(d), but it has a much lower amplitude than the unphysical wave and does not generate visible features in Fig.~\ref{fig6}(c). The simulated frames from Supplementary Movie 3 were calculated using the velocity-capped CFD model.

The image simulations also clarify that the size of the CFD-modeled ablation core is larger than in the experiment. This would be difficult to confirm without the image simulations from Fig.~\ref{fig4}, because the drop magnifies and distorts the core image. A known limitation of CFD models of ablation is approximating the formation of plasma and the phase transitions in the laser-heated region \cite{linz}, but it was not clear how this impacts the ablation dynamics. Our image simulations show that this limitation impacts the core size, even when the CFD model replicates the waveform of the shock at the drop surface. Thus, future CFD models should replicate both the waveform of the shock at the surface and the size of the ablation core.

More generally, high-fidelity image simulations can be used to test hypotheses about systems that are not well understood. For example, if an image of a fluid object displays an unexpected dark region, image simulations can test if this region is due to a bubble out of focus, due to a pressure wave, or is simply a blurred image of another component in the experimental setup.

\subsection{\label{sec3B}Detection of axial misalignments}

In the drop ablation experiments the X-ray pulse should pass through the center of the drop, but vibrations due to vacuum pumps and drop breakup instabilities can cause micron-scale misalignments, which change the pressure distribution of the reflected negative pressure wave. Misalignments are difficult to detect if they occur along the direction of imaging, but the image of the negative pressure wave is sensitive to them. For the ablation movie that generated an ablation hole, we initially used an axisymmetric CFD model, and the corresponding simulated images did not reproduce the vertical-to-horizontal aspect ratio of the negative pressure wave.

\begin{figure}
\includegraphics{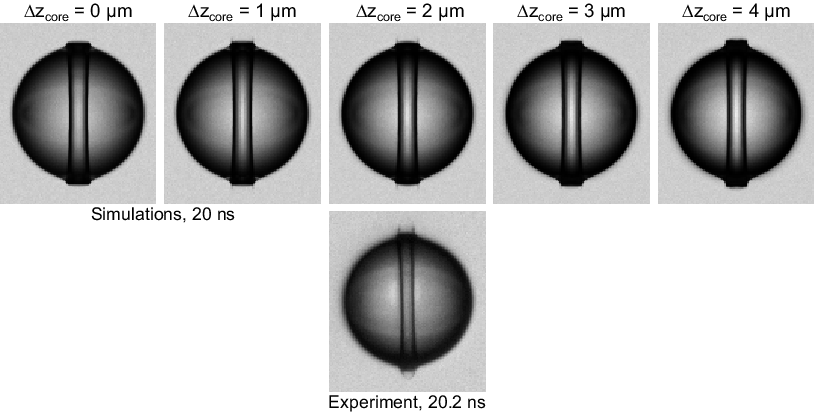}
\caption{\label{fig7}Determination of the misalignment between the drop and X rays. The experimental image of a negative image is compared to simulations made for a range of misalignments. The aspect ratio of the negative pressure wave is best replicated at 2–3 µm. Differences between the simulated images may only be visible on a computer screen. }
\end{figure}

To determine the experimental misalignment, we performed non-axisymmetric CFD simulations where the core axis was displaced, along the imaging direction, from the center of the drop. Figure~\ref{fig7} shows the experimental image at 23 ns from Fig.~\ref{fig4}, along with corresponding simulated images for core misalignments ranging from 0 µm to 4 µm (in the direction of light propagation). The best match, based on visual comparisons of the aspect ratio, was obtained with the core misaligned by 2–3 $\mu$m. Thus, high fidelity image simulations allowed us to discover and quantify experimental imperfections that are difficult to detect because they occur along the imaging direction.

\subsection{\label{sec3C}Sensitivity of images to the amplitude of pressure waves}

In both experimental and simulated images of the ablated drops, the contrast of the shock wave increases with the amplitude of the pressure waves. We evaluated the sensitivity of the images to the wave pressure by performing image simulations of waves with different amplitudes. As a reference case, we used the 5 ns CFD simulation for the weak ablation experiment, where the core did not form (Supplementary Movie 3). To generate drop models with different pressure amplitudes, we scaled linearly the difference between the densities of the ablated drop and of a spherical drop, such that the peak density increased by a fixed amount. To detect small changes in the images, we integrated a horizontal slice in the image of the ablated drop and subtracted it from the corresponding slice in the image of a spherical drop.

\begin{figure}
\includegraphics{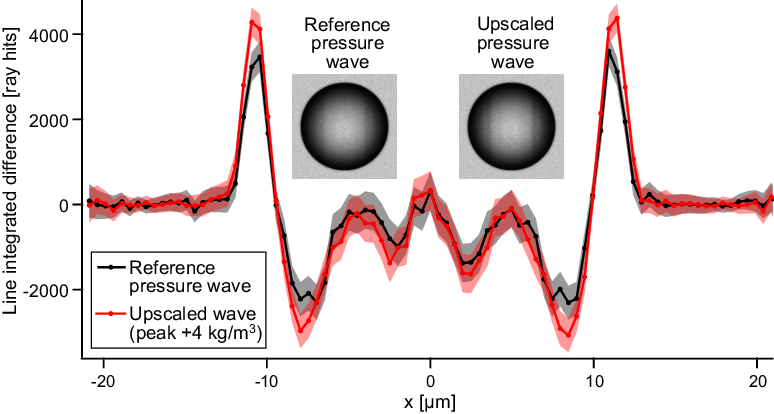}
\caption{\label{fig8}Sensitivity of simulated images to wave pressures. The graph shows integrated difference profiles of image simulations for the weak ablation experiment, for the reference simulation and for another simulation with a linearly upscaled density distribution. For a density distribution with a peak increase of 4 kg/m$^3$, the peaks of the profiles differ more than the shot noise, which is indicated with color bands at $\pm$1 standard deviation. }
\end{figure}

Figure~\ref{fig8} shows the line integrated image differences for the reference pressure wave, and for a density-upscaled wave. The bands around the profiles represent the standard deviation of the data due to shot noise, calculated as the square root of the sum of ray hits in the images of the ablated drop and the spherical drop. The profile of a wave with peak density upscaled by 4 kg/m$^3$, corresponding to a pressure variation of 8 MPa, was significantly different from the profile of the reference wave. Since the experimental images had only 30–40\% larger noise than the simulations, we estimate that the experimental images were sensitive to pressure changes around 10 MPa. It should thus be possible to measure the pressure amplitude with an accuracy around 10 MPa, by comparing experimental and simulated profiles. For the present data, this was not possible because the experimental drop was misaligned axially while the simulated drop used a surrogate light field source, which led to differently shaped profiles for experiment and simulation.

The peak density change of 4 kg/m$^3$ corresponds to changes in the projected mass density below $10^{-5}$ kg/m$^2$, orders of magnitude smaller than the mass density differences observed in XPCI \cite{hodge}. This sensitivity can be improved by reducing the shot noise via using higher illumination intensities in the experiment. For the experimental setup we used, the illumination intensity can be improved by at least one order of magnitude by using higher energy laser pulses, a tighter illumination spot, and a higher magnification to prevent the saturation of the camera. With an order of magnitude higher illumination intensity, peak density changes around 1 kg/m$^3$ should be detectable.

\subsection{\label{sec3D}Extraction of 3D object properties via optimization}

The ability to reproduce images of 3D objects suggests that ray-traced image simulations can be used to extract 3D properties from single 2D images. For example, the good agreement between simulation and experiment for the focus scan in Fig.~\ref{fig2} indicates that an unknown axial position of a drop can be determined by finding the best matched image in a series of simulations at different axial positions. An estimate within a few microns can be done with visual comparisons, both for the axial position of a spherical drop (Fig.~\ref{fig2}) and the axial misalignment of the laser-ablated drop (Fig.~\ref{fig7}). Using numerical optimization, a better precision can be achieved and additional parameters can be determined. Optimization is used in other imaging methods, such as model-based optical holography \cite{martin}, but optimization with ray-traced images requires a specialized framework that includes fast noisy simulations, image similarly measures, and stochastic optimization.

Optimization requires a numerical quantification of the closeness of experimental and simulated images, which defines an error function (or objective function) whose value can then be maximized or minimized. The choice of an error function for ray traced images is related to the problem of quantifying the similarity of images (Section II.C). Computing the mean-squared difference or error between images (MSE) is the simplest approach, but other measures such as structural similarity indices \cite{wangSSIM,wangMSSIM,sampat} may perform better.

A specific challenge in ray tracing is the long computation time needed to replicate an image with the same noise as the experimental ones (Sec.~\ref{sec2D}). Optimization involves generating and testing many trial images, and doing so with the simulations shown in Figs.~\ref{fig2}–\ref{fig4} would be impractical. Fortunately, as shown in Fig.~\ref{fig5}, key image features can be preserved in simulations using orders of magnitude fewer rays, and proportionally faster computations.

\begin{figure}
\includegraphics{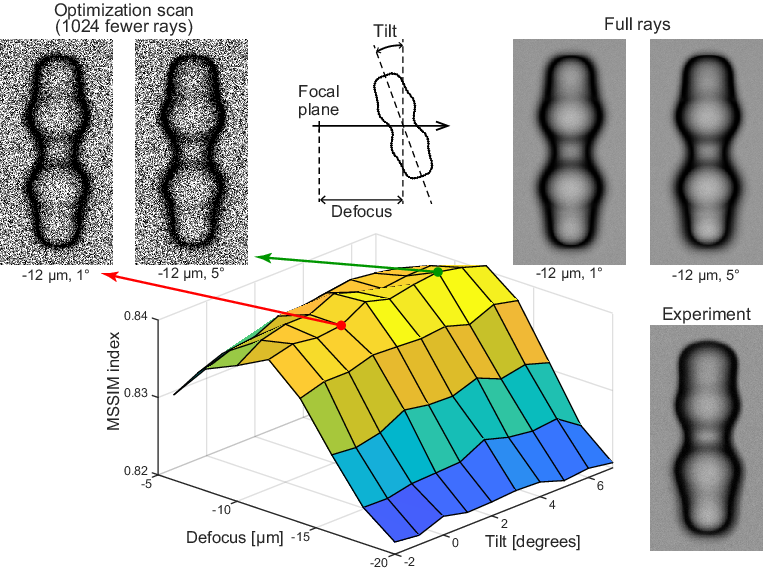}
\caption{\label{fig9}Rapid extraction of 3D object properties with noisy simulations. The injector was tilted to generate merging drops with defocus and tilt, whose values were optimized using a parametric grid scan. The location of the best multiscale structural similarity index (MSSIM \cite{wangMSSIM}) between the experiment and simulations remained stable as the number of rays was reduced to 1/1024 of the number of experimentally recorded photons. }
\end{figure}

As an example of efficiently extracting multiple 3D parameters, Fig.~\ref{fig9} displays a grid-scan optimization of two depth parameters of the merging drops (defocus and tilt along imaging direction). To explore how much the optimization can be sped up, we reduced the number of rays in factors of 2 while monitoring the behavior of three error functions equal to the similarity measures MSE, MSSIM and CWSSIM (see Sec.~\ref{sec2C}). As the number of rays was reduced, the location of the optimum drifted away, and the optimization surface became noisy due to the increased shot noise of the simulations \cite{suppmat}. The degree of degradation varied between the error functions, and MSSIM allowed the furthest reduction, to 1/1024 of the initial number of rays. Using MSSIM and 1024 fewer rays, the entire grid scan in Fig.~\ref{fig9} could be completed with one order of magnitude fewer rays than a single image from Fig.~\ref{fig3}.

For optimization problems with more parameters, optimization algorithms can be used. The stochastic noise of ray-traced images leads to noisy error functions, and this prevents the use of gradient-descent optimization algorithms. Instead, stochastic physics models can be optimized using direct search algorithms \cite{kolda}. For example, Kalita \emph{et al.} \cite{kalita} used a pattern search algorithm to determine the parameters of a Monte Carlo model for the freezing of supercooled water drops. We used this algorithm, along with similarity measures as error functions, to optimize simultaneously the five optical parameters of a sphere: 3D position, radius, and refractive index. In model-based holography, this problem is a showcase for 3D-quantitative imaging \cite{martin,lee}. For the droplet at $-$10 µm from Fig.~\ref{fig2}, the five-parameter optimization produced a single minimum close to the expected values, with an estimated uncertainty of 100 nm for the lateral position, 400 nm for the axial position, 50 nm for the radius, and 0.005 for the refractive index \cite{suppmat}. Compared to the precision that can achieved with holographic imaging \cite{lee}, the absolute uncertainties of positions are worse, and the relative uncertainties of the radius and of the refractive index are comparable. To our knowledge, this is the first time these parameters were extracted from a basic optical image of an out-of-focus transparent sphere.

Improving the detail of optimized objects requires more complex models, but the volume of the search space for optimization scales exponentially with the number of parameters. The search volume can be reduced by using object models that provide a more accurate starting point for optimization. Thus, instead of using ray traced images directly for optimization, they could first be used qualitatively, to refine other simulations that generate the initial object. High fidelity ray-tracing image simulations are well suited for this strategy because they can handle objects with complicated optical properties, and because they produce images that can be interpreted visually. For example, in Sec.~\ref{sec3A}, we showed how image simulations can be used to identify and correct missing features in CFD ablation models. This combination of high-fidelity image simulations and CFD promises to increase the modeling accuracy of laser ablation and cavitation, and opens a path towards measuring 3D properties such as density and pressure from optical images.

The combined use of ray-tracing simulations and of an independent model can also prevent overfitting the experimental images, in the sense of improving precision without improving accuracy. In principle, the amplitude of pressure waves can be adjusted to improve the match of simulations in Fig.~\ref{fig4}, but this is not guaranteed to improve the accuracy of the pressure, because the CFD model has at least one unsolved deficiency (see Sec.~\ref{sec3A}). Here, to improve the accuracy of characterizing the X-ray laser ablation in drops, the best next step is to improve the CFD model.

\section{\label{sec4}Conclusion}

Ray tracing is one of the oldest optical techniques, first documented more than 1000 years ago in Ibn Sahl’s treatise on optics \cite{zghal}. For such a basic and well understood technique, the lack of fidelity in images generated with complex ray-tracing models may seem surprising. Nevertheless, despite their complexity, previous ray-tracing models were incomplete. Here, we developed high fidelity image simulations based on ray tracing, which required two new techniques: using a light field source, and testing the fidelity of the simulations via focus scans. After they were implemented, we found that the image simulations maintained high fidelity across a set of objects that tested all the basic ways in which light interacts with transparent objects. We therefore expect that implementing these two techniques will lead to high fidelity image simulations, at least for brightfield microscopic imaging and when using Zemax for ray tracing. Since our findings are applicable to microscopy, they are applicable in other fields, such as microfluidics or biology.

The ray-traced image simulations can be used not only in future studies, but also with previously recorded data, if the optical elements can be modeled and the light field can be measured. For images recorded with commercial microscopes, which have a fixed setup, accurate optics modeling and light field measurements should be feasible. Here, we used this approach for the laser ablated drops, and we could replicate images of pressure waves despite lacking an experimentally calibrated light field.

The physical fidelity of image simulations is important because it enables both qualitative and quantitative applications. As qualitative applications, we have shown that ray-traced images can be used to discover experimental imperfections and missing features in other physical models. More generally, the simulations can test hypotheses about unexplained features in images. These qualitative applications benefit from the visual interpretability of basic imaging and can help understand new phenomena that cannot be quantified yet. As quantitative applications, we showed that 3D properties can be retrieved by optimizing ray-traced images or constrained by combining ray-traced images with CFD. Although the retrieval of 3D properties can be done with other techniques, using basic imaging is experimentally simpler, and avoids the loss of resolution in plenoptic imaging or the loss of visual interpretability in holographic imaging. Since the retrieval of 3D properties with ray traced images is a new approach, there are multiple opportunities for improvement, including multi-focal or multi-perspective imaging to constrain tighter the optimization, faster ray tracing algorithms, and more efficient optimization with noise-tolerant argorithms or noise-tolerant similarity measures.

\begin{acknowledgments}
\emph{Acknowledgments}—This work was primarily supported by the National Science Foundation under Grant No. 2123634. Supplement funding for this project was provided by the Rutgers University–Newark Chancellor’s Research Office. A.K. acknowledges support from the Rutgers University, Graduate School-Newark Dean's Dissertation Fellowship. N.A.A. acknowledges funding by the European Research Council (ERC) under the Advanced Grant Project No. 101094463 (GENUFASD). T.P., A.B., S.A. and N.A.A. gratefully acknowledge the Gauss Centre for Supercomputing e.V. for funding this project by providing computing time at Leibniz Supercomputing Centre. Use of the Linac Coherent Light Source (LCLS), SLAC National Accelerator Laboratory, was supported by the US Department of Energy, Office of Science, Office of Basic Energy Sciences under Contract no. DE-AC02-76SF00515.

\emph{Data availability}—A data and software repository for running simulations from Figs. 2-4 is available at DOI: 10.5281/zenodo.17772690 \cite{kalitaZD}. Other data that supports the findings of this manuscript, and were not included in the main text or the Supplemental Material, are available from the authors upon reasonable request.

\emph{Note}—The manuscript has Supplmentary Materials (videos and text) and includes Refs. \cite{stanZD, vargaftik}.
\end{acknowledgments}

\end{document}